\documentclass[useAMS,usenatbib]{mn2e}

\usepackage{psfig}

\title[A tilted disk in GR\,Mus]{The origin of the tilted disk in the 
low mass X-ray binary GR\,Mus (XB\,1254$-$690)}

\author[R. Cornelisse et~al.]{R. Cornelisse$^{1,2}$\thanks{E-mail:
    corneli@iac.es}, M.M. Kotze$^{3,4}$, J. Casares$^{1,2}$, P.A. Charles$^{5,4}$, P.J. Hakala$^{6}$\\
  $^{1}$ Instituto de Astrofisica de Canarias, Via Lactea, La Laguna, 
E-38200, Santa  Cruz de Tenerife, Spain\\
  $^2$ Departamento de Astrofisica, Universidad de La Laguna, 
E-38205, La Laguna, Tenerife, Spain\\
  $^3$ South Africa Astronomical Observatory, P.O.Box 9, Observatory 7935, 
South Africa\\
 $^4$ Astrophysics, Cosmology and Gravity Centre (ACGC), Astronomy Department,
University of Cape Town, Rondebosch 7701, South Africa\\
 $^5$ School of Physics and Astronomy, University of Southampton, 
Highfield, Southampton SO17 1BJ, UK\\
 $^6$ Finnish Centre for Astronomy with ESO (FINCA), V\"ais\"al\"antie 20, University of Turku, FIN-21500, Piikki\"o\\
}
\begin{document}

\date{Accepted Received; in original form}

\pagerange{\pageref{firstpage}--\pageref{lastpage}} \pubyear{2013}

\maketitle

\label{firstpage}

\begin{abstract}
We present photometric and spectroscopic observations of the low mass
X-ray binary GR\,Mus (XB\,1254$-$690), and find strong evidence for
the presence of a negative superhump with a period that is
2.4$\pm$0.3\% shorter than the orbital. This provides further support
that GR\,Mus indeed harbours a precessing accretion disk (with a
period of 6.74$\pm$0.07 day) that has retrograde precession and is
completely tilted out of the orbital plane along its line of
nodes. This tilt causes a large fraction of the gas in the accretion
stream to either over- or underflow the accretion disk instead of
hitting the disk rim, and could be a feature of all low mass X-ray
binaries with characteristics similar to GR\,Mus (i.e. the so-called
atoll sources). Furthermore, we also find marginal evidence for the
presence of a positive superhump, suggesting that the accretion disk
  in GR\,Mus is eccentric due to tidal resonances. If true, than
    the relationship between the positive superhump period excess and
    the mass ratio ($q$) provides a constraint of $q$=$M_{\rm
      donor}$/$M_{\rm NS}$=0.33-0.36.  Together with the radial
    velocity semi-amplitude measurements of the compact object, and
    previous modeling of the inclination we obtain a mass for the
    neutron star of 1.2$\le$$M_{\rm NS}$/$M_\odot$$\le$1.8 (95\%
    confidence).
\end{abstract}


\section{Introduction}

For about a dozen persistently active low mass X-ray binaries
(LMXBs), i.e. binaries that harbour a compact object (either a black
hole or a neutron star) and a low mass ($\simeq$1$M_\odot$) companion,
the optical counterpart is known (e.g. Charles \& Coe 2006). Despite
being relatively bright optically, kinematic studies to constrain the
mass of these systems have been difficult. In general, the reprocessed
X-ray emission in the outer accretion disk, that is formed around the
compact object, dominates the optical flux thereby swamping any
intrinsic spectral feature of the donor star. However, spectroscopic
signatures of the irradiated face of the donor star have now been
detected in many LMXBs in the form of narrow emission components
(e.g. Steeghs \& Casares 2002). This signature is most visible in the
Bowen region, a blend of N\,III 4634/4640\AA\, and C\,III 4647/4650
\AA\, lines, and is for many optically bright LMXBs the only way to
constrain the mass function thus far (see e.g. Cornelisse et~al. 2008
for an overview).

One such LMXB for which a signature of the irradiated surface of the
donor star in the Bowen region is present is XB\,1254$-$690 (Barnes
et~al. 2007). Its optical counterpart was identified with the
$V$$\simeq$19 blue star GR\,Mus (Griffiths et~al. 1978), and its
Type\,I X-ray bursts identify the compact object as a neutron star
(Mason et~al. 1980). Interestingly, GR\,Mus also shows periodic X-ray
dips with a recurrence time of $\simeq$3.9 hrs (Courvoisier
et~al. 1986), which is similar to the optical modulation detected by
Motch et~al. (1987). These dips are thought to be due to periodic
obscuration of the inner accretion disk by a structure located in the
outer regions of the disk (White \& Swank 1982), and not only
  suggest that this $\simeq$3.9 hr period is orbital but also that
GR\,Mus has a reasonably high inclination. For example, modeling of the
optical lightcurves by Motch et~al. (1987) finds that the
inclination of GR\,Mus must be between 65$^{\circ}$ and 73$^{\circ}$.

\begin{figure*}
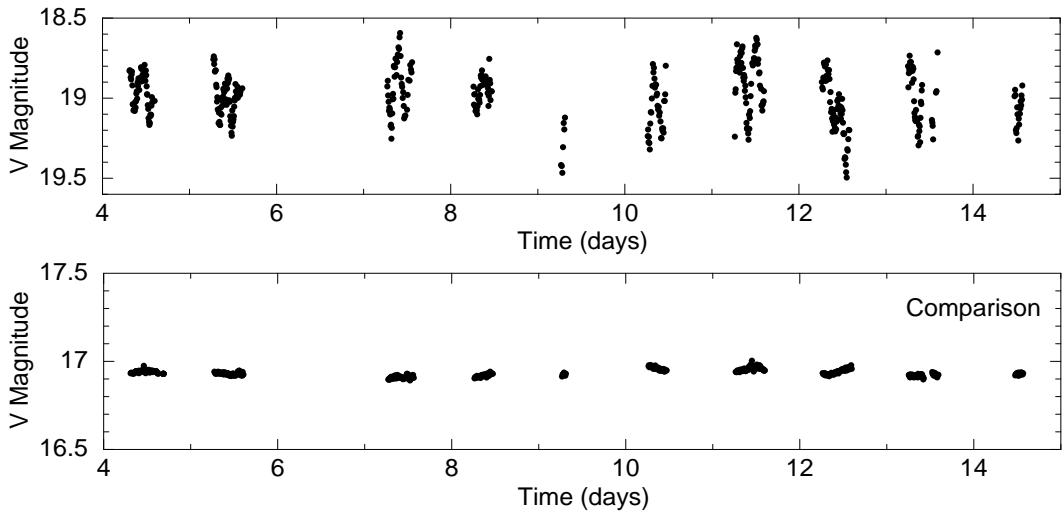
\begin{center}
\psfig{figure=total_V.ps,angle=-90,width=14cm}
\psfig{figure=comp_V.ps,angle=-90,width=14cm}
\caption{$V$-band light curve of our 11 day campaign on GR\,Mus
  (top). For illustration purposes we have also included a
  comparison star (bottom) to show that the variability observed in
  GR\,Mus is intrinsic.
\label{light}}
\end{center}\end{figure*}

Compared to most other dipping LMXBs such as XB\,1916$-$053
(e.g. Smale et~al. 1992), the large variation in both dip depth and
length in GR\,Mus is unusual. For example, dips with a reduction in
the 1-10 keV X-ray flux of $\simeq$95\% and ``only'' $\simeq$20\% were
observed by Courvoisier et~al. (1987). On the other hand, Smale \&
Wachter (1999) and Iaria et~al. (2001) found no dips with an
upper-limit of $\simeq$2\%. Further observations showed stages without
any dipping, only shallow dipping or strong dipping again (e.g. Smale
et~al. 2002; Diaz-Trigo et~al. 2009). A characteristic timescale of
$\sim$60 hrs for the recurrence of dipping behavior was found by
Diaz-Trigo et~al. (2009), and they also suggested that the general
dipping behavior could be explained by the presence of a precessing
accretion disk that is tilted out of the orbital plane.

In this paper we present the results of a 2 week photometric campaign
combined with 2 orbital periods of spectroscopic observations on
GR\,Mus. We find strong evidence for the presence of a negative
superhump, strongly suggesting that GR\,Mus indeed harbours a
tilted accretion disk which is the origin of the large variation in
the properties of the X-ray dips. In Sect.\,2 we present an overview
of our observations. In Sect.\,3 we show the results of our
photometric campaign, while Sect.\,4 discusses the results of our
spectroscopic observations. We continue with a discussion on the
evidence for a tilted disk in Sect.\,5.1 and provide further
constraints on the system parameters of GR\,Mus in Sect.\,5.2. We
finish with a brief conclusion in Sect.\,6.

\section{Observations and Data Reduction}

\subsection{Photometric observations}

From 17-27 April 2010 we obtained 11 nights of photometric
observations of GR\,Mus using the 1.0m telescope at SAAO. We used the
SAAO STE4 CCD camera, which is a SITe back-illuminated detector with
1024$\times$1024 pixels. During the first night we obtained alternate
$V$ and $I$ band images of GR\,Mus, while during subsequent nights we
alternated between $V$ and $R$. The integration time for each image
was always 180 sec. Since the majority of our observations were in
$V$, and the resulting lightcurves in the different bands show a very
similar morphology, we will concentrate only on this band for the rest
of the paper. In a future paper we plan to present modeling of the
resulting lightcurves for all bands. In Table\,\ref{log} we give an
overview of our $V$-band observations. Excluding observations for
which the weather conditions were too poor, we obtained a total of 448
images spread over 10 nights.

\begin{table}\begin{center}
\caption{GR\,Mus Observing Log.
\label{log}}
\begin{tabular}{cccc}
\hline
Date/start time & \# obs. & exp. time & seeing \\
(dd-mm-yy/UT) &  & (s) & (arcsec)\\
\hline
{\bf Photometry}  \\
17-04-10/19:00  & 56 & 180 & 1.2-1.4\\
18-04-10/18:20  & 67 & 180 & 1.1-1.5\\
19-04-10/closed & 00 & -- & --\\
20-04-10/18:20  & 47 & 180 & 1.0-2.0\\
21-04-10/18:00  & 45 & 180 & 1.1-2.5\\
22-04-10/18:15  & 07 & 180 & 2.0-2.5\\
23-04-10/18:10  & 36 & 180 & 1.3-2.5\\
24-04-10/18:10  & 66 & 180 & 1.2-1.6\\
25-04-10/18:10  & 62 & 180 & 1.0-1.4\\
26-04-10/18:10  & 43 & 180 & 1.1-2.0\\
27-04-10/23:15  & 19 & 180 & 1.0-2.5\\
\hline
{\bf Spectroscopy} \\
16-04-12/03:58 & 22 & 660 & 0.5-1.6\\
20-04-12/02:20 & 25 & 660 & 0.6-1.0\\
\hline

\end{tabular}
\end{center}\end{table}

For the reduction we used a point-spread function (PSF) fitting
program based on the DoPhot routine by Schechter et~al. (1993). This
routine is specifically designed for bias subtraction, and source
extraction (using both aperture photometry and PSF-fitting) for the
STE4 CCD on the 1.0m telescope at SAAO. We extracted the counts for
GR\,Mus and 3 comparison stars so as to perform differential photometry. The
magnitude of each target was obtained using a standard star that was
observed each night.

\subsection{Spectroscopic observations}

\begin{figure*}[t]\begin{center}
\parbox{4.0cm}{\psfig{figure=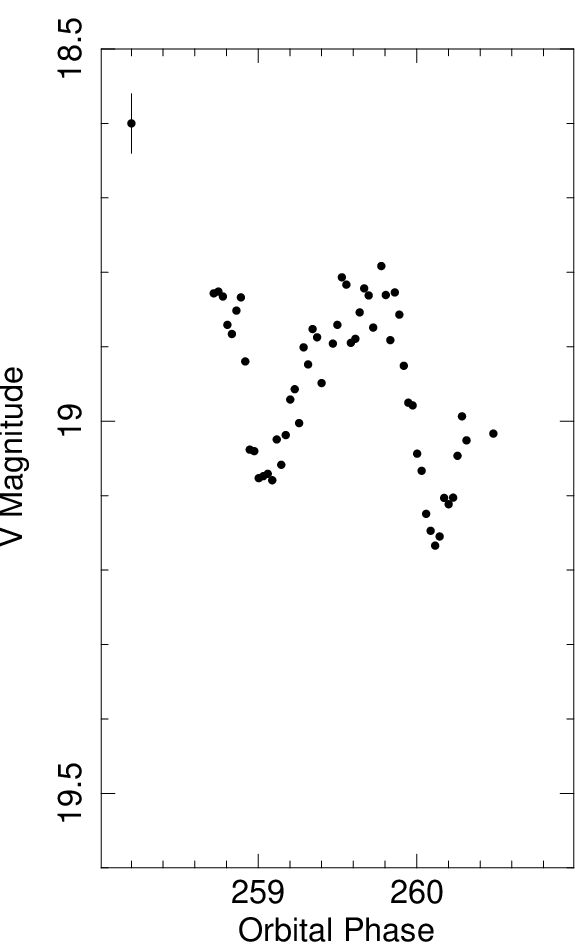,angle=-90,width=4cm}}
\parbox{4.0cm}{\psfig{figure=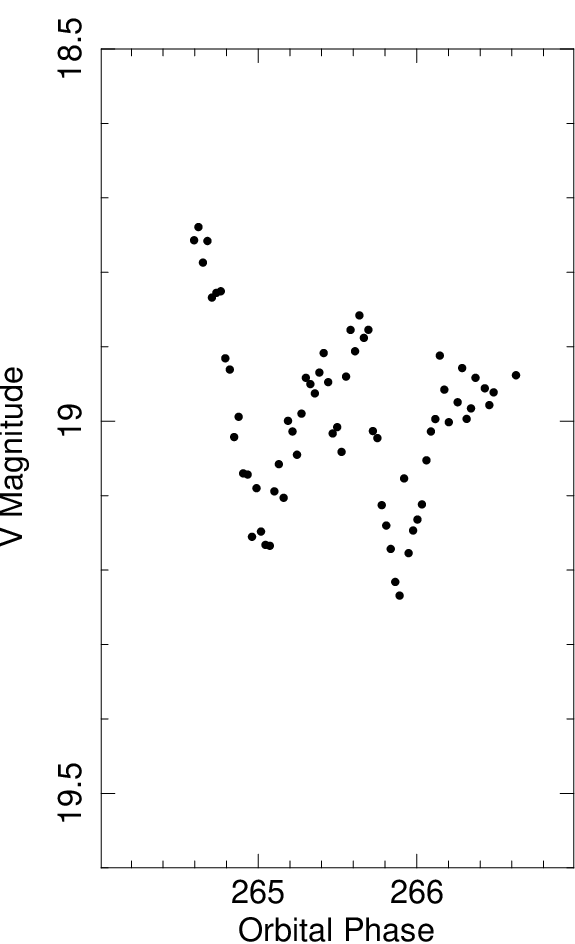,angle=-90,width=4cm}}
\parbox{4.0cm}{\psfig{figure=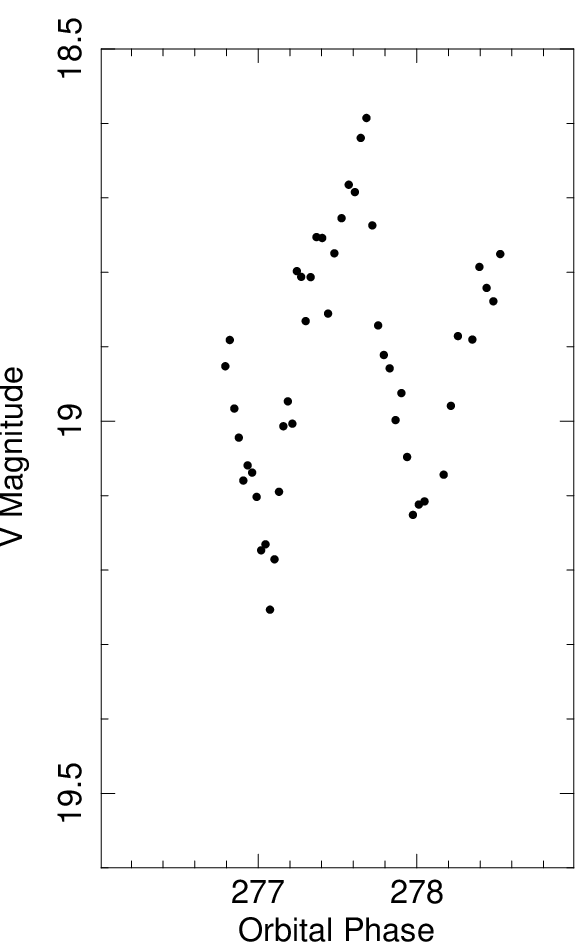,angle=-90,width=4cm}}
\parbox{4.0cm}{\psfig{figure=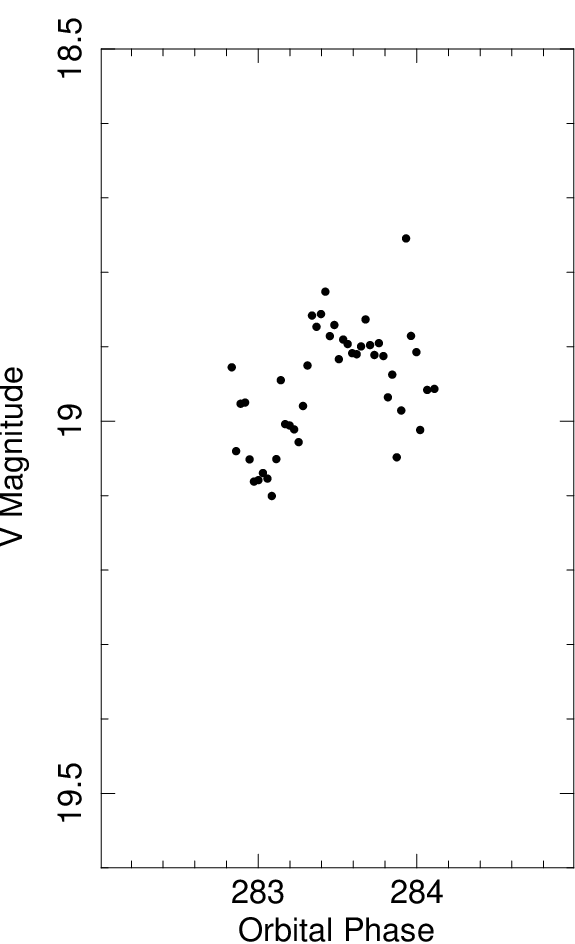,angle=-90,width=4cm}}
\parbox{4.0cm}{\psfig{figure=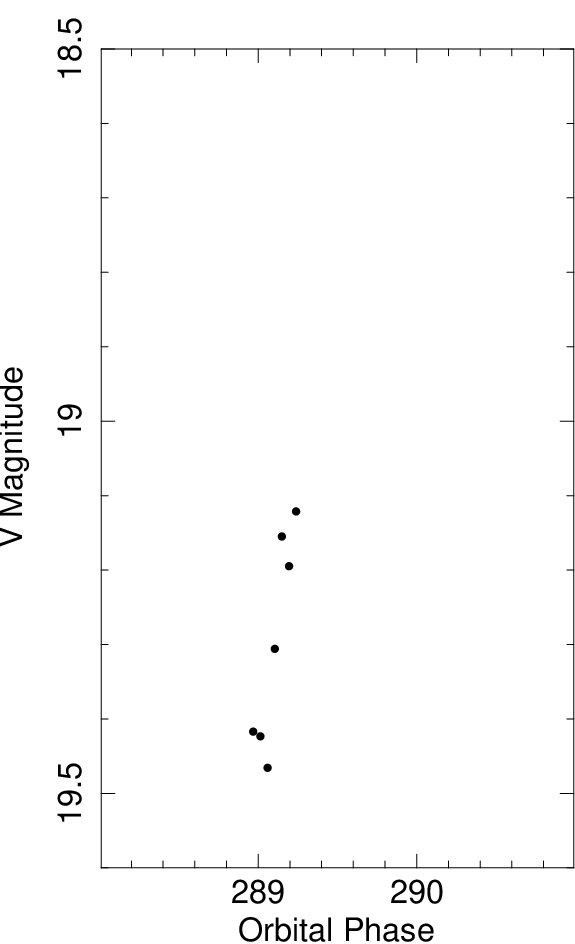,angle=-90,width=4cm}}
\parbox{4.0cm}{\psfig{figure=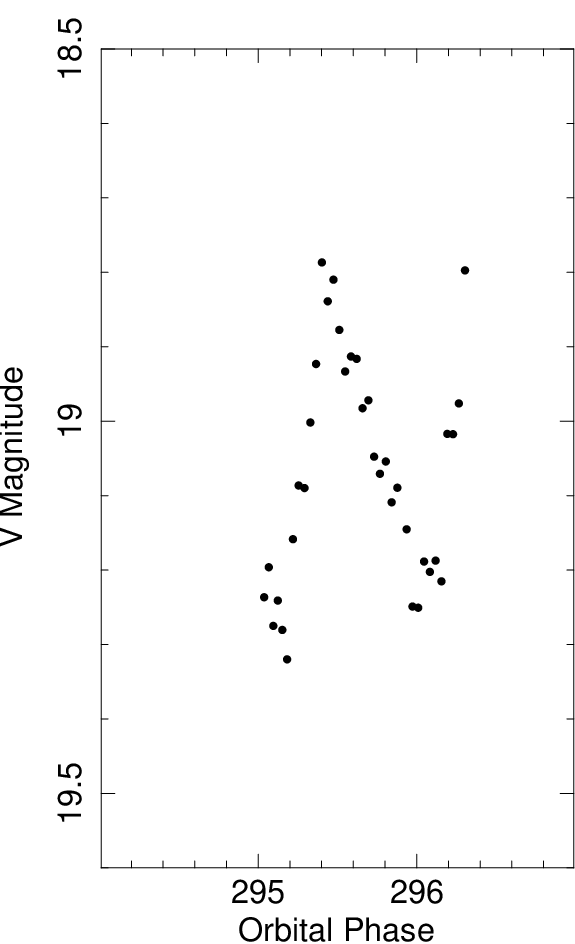,angle=-90,width=4cm}}
\parbox{4.0cm}{\psfig{figure=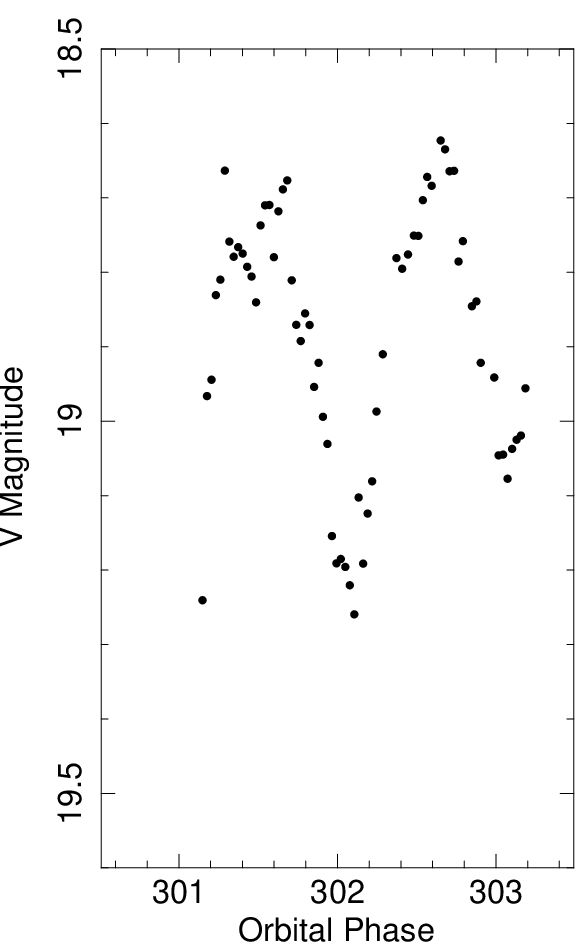,angle=-90,width=4cm}}
\parbox{4.0cm}{\psfig{figure=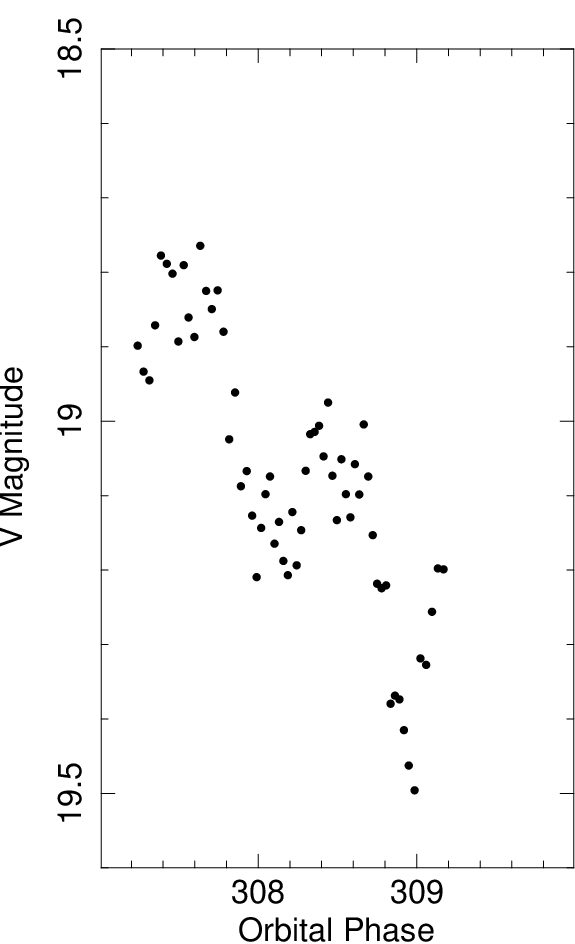,angle=-90,width=4cm}}
\parbox{4.0cm}{\psfig{figure=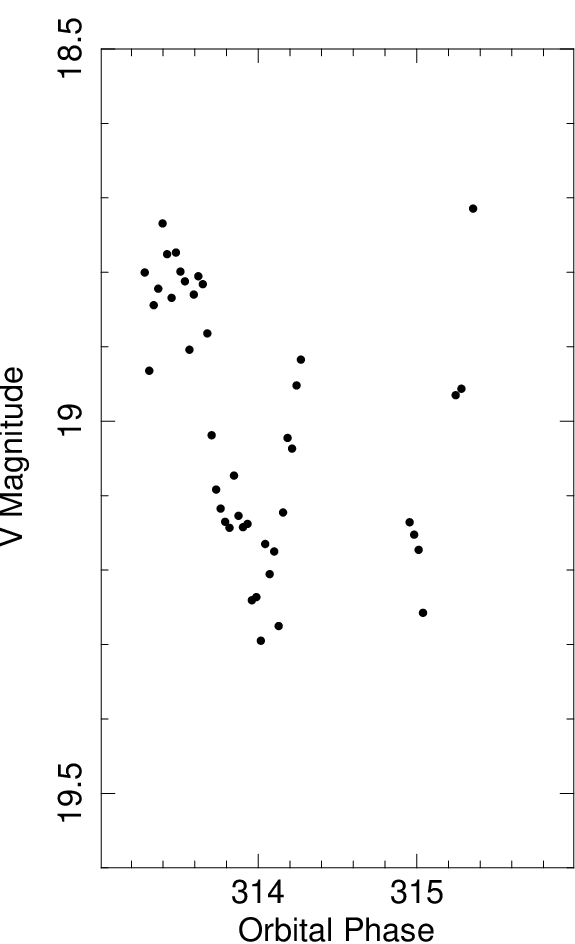,angle=-90,width=4cm}}
\parbox{4.0cm}{\psfig{figure=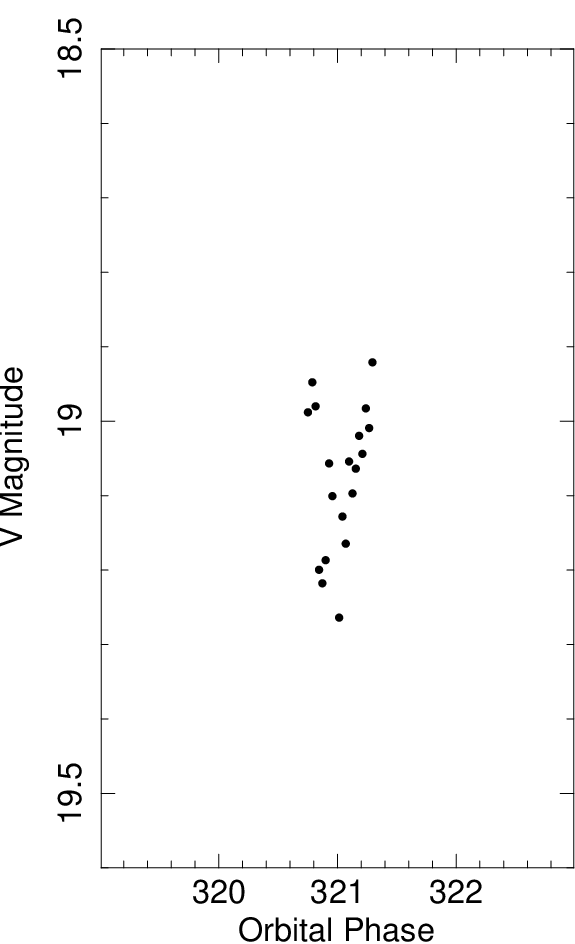,angle=-90,width=4cm}}
\caption{Close-up lightcurves of each individual night during our
  photometric campaign on GR\,Mus. We have converted the time of each
  observation into the orbital cycle using the ephemeris of Diaz-Trigo
  et~al. (2009). A typical error bar is indicated in the top-left box.
\label{stamps}}
\end{center}\end{figure*}

On both April 16 and 20 2012 we obtained approximately one orbit of
phase-resolved spectroscopy of GR\,Mus using the FORS\,2 spectrograph
attached to the VLT Unit 1 at Paranal Observatory (ESO). During each
night we used the 1200g volume-phased holographic grism with an
integration time of 11 min each. We used a slit width of 0.7$''$,
giving a wavelength coverage of $\lambda$$\lambda$4087-5561 with a
resolution of 106 km s$^{-1}$ (FWHM). We aligned the slit in such a
way to avoid any interference from nearby brighter stars. This does
mean that no comparison star is included in the slit and we could
therefore not correct for slit losses. Arc lamp exposures were taken
for wavelength calibration during the daytime. In Table\,\ref{log} we
give an overview of the observations.

\begin{figure}
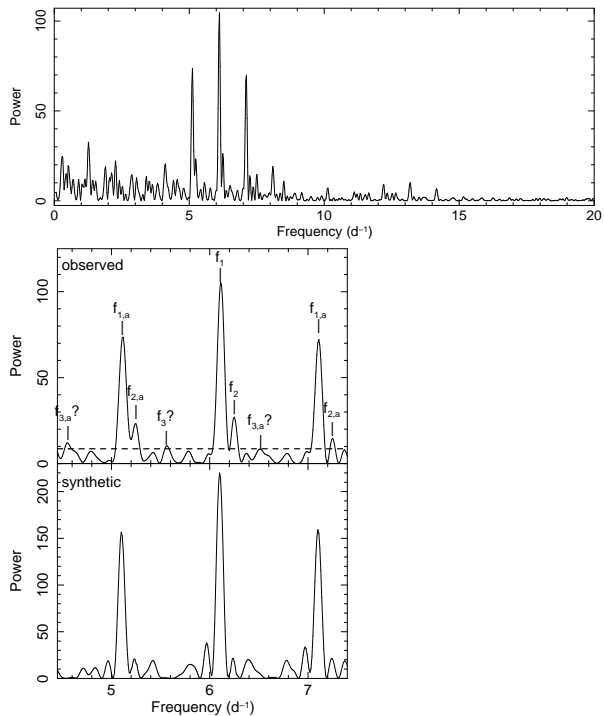

\centering \psfig{figure=scar_all.ps,angle=-90,width=8cm}
\centering \psfig{figure=scar_V.ps,angle=-90,width=4.5cm}
\caption{{\it Top:} Lomb-Scargle periodogram of our photometric
  campaign on GR\,Mus showing the full frequency range. {\it Bottom:}
  A close-up of the interesting region of the periodogram (top) and
  the periodogram of simulated data (bottom; see text). We have
  indicated the 3 most prominent peaks with $f_X$ (where $X$ is 1,2 or
  3), and their corresponding daily aliases (labelled as
  $f_{X,a}$). Furthermore, the dashed line indicates the 3$\sigma$
  confidence level. The marginally significant period $f_3$ has been
  labelled with ``?''.
\label{scar}}
\end{figure}

The PAMELA software was used for the data reduction (i.e. debias,
flatfielding etc), which allows for optimal extraction of the spectra
(Horne 1986). The pixel-to-wavelength scale was determined using a 4th
order polynomial fit to 8 lines resulting in a dispersion of 0.73
\AA\,pixel$^{-1}$, an rms scatter of $<$0.01 \AA, and a wavelength
coverage of $\lambda$$\lambda$4434-6060. The final resolution around
the He\,II $\lambda$4686 emission line is 107.6 km s$^{-1}$
(FWHM). Finally we also corrected for any velocity shifts due to
instrument flexure (always $<$10 km s$^{-1}$) by cross-correlating the
sky spectra. For the corresponding analysis of the resulting dataset
we used the MOLLY package.

\section{Photometry}

In Fig.\,\ref{light} we present the resulting lightcurve of our 2
weeks of $V$ band photometric observations on GR\,Mus. As a check we
also include the lightcurves of a comparison star, in which no
strong variability is present, and therefore all variability in
GR\,Mus is intrinsic and not due to weather or instrumental
variations. 

The first thing to note is that the average magnitude ($V$$\simeq$19)
is similar to previous observations of GR\,Mus (e.g. Motch
et~al. 1987; Smale \& Wachter 1999). Furthermore we note that there is
not only strong variability on a nightly timescale (with an amplitude
of $\simeq$0.5 magnitude), but also substantial changes from night to
night and even over a longer period. This strongly suggests that
something is changing systematically in GR\,Mus on a timescale much
longer than the orbital period. Finally we note the presence of two
``dip''-like events in Fig.\,\ref{light} about 3.3 days apart (around
9.3 and 12.6 days).

In order to investigate the origin of the nightly variability, we
converted the time into orbital cycles using the ephemeris by
Diaz-Trigo et~al. (2009). In Fig.\,\ref{stamps} we show ``postage
stamps'' of the individual nights. We note that the nightly variation
in GR\,Mus is mostly periodic with the minima in the lightcurve
corresponding to orbital phase zero of the ephemeris of
Diaz-Trigo. However, there are nights (e.g. around orbital cycles
307-309) where the variations cannot be explained solely by a
$\simeq$4 hour periodic variability, but where a longer-term trend is
also present that gives rise to a large variation from cycle to
cycle. Furthermore, we also note that the amplitude of the variability
changes from $\simeq$0.2 (cycle 283) to $\simeq$0.7 magnitude
(e.g. cycles 301-303) between different cycles. In total this strongly
suggests that, just like the X-ray emission (Diaz-Trigo et~al. 2009),
the optical emission cannot be described by simple orbital motion
only.

Since Fig.\,\ref{light} implies the presence of a longer-term trend in
GR\,Mus, we therefore determined, where possible, the minimum
magnitude for each cycle (as Fig.\,\ref{stamps} shows that they are
better defined than the maximum). Furthermore, for each night we
determined the semi-amplitude of the $\simeq$4 hr periodic variability
by fitting simultaneously a sine curve and a linear trend. However, no
obvious trends are visible in both minimum and semi-amplitude as a
function of cycle, nor is there any trend when plotting the minimum as
a function of semi-amplitude.

\begin{figure*}\begin{center}
\psfig{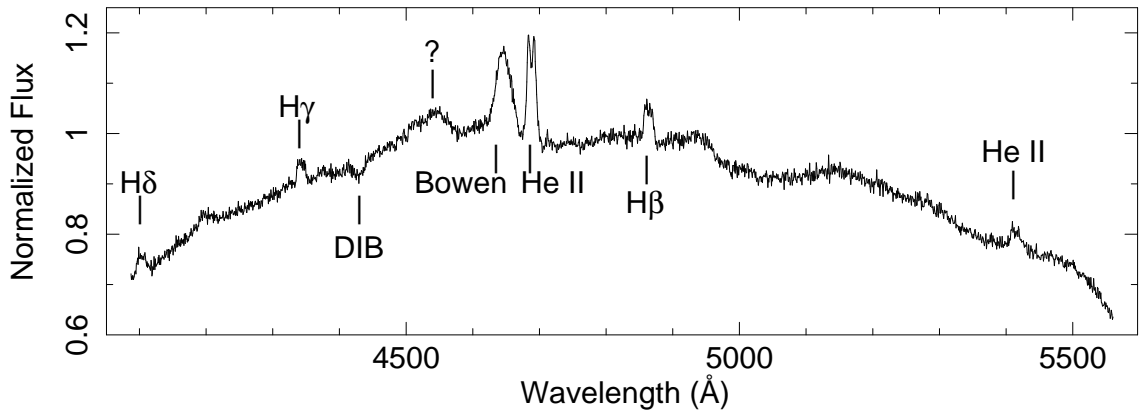}
\caption{Average spectrum of GR\,Mus from our 2 orbits of VLT data
  normalized on the continuum level around He\,II $\lambda$4686. We
  have indicated the most prominent lines. The diffuse interstellar
  line at 4430 \AA\, is indicated with ``DIB'', while the unknown
    feature at 4530 \AA\ is indicated with ``?''.
\label{average}}
\end{center}\end{figure*}

Another way to obtain more information about the origin of the extra
variability in GR\,Mus is by searching for periodic variation using
the Lomb-Scargle technique (Scargle 1982). In Fig.\,\ref{scar}/top we
present the resulting periodogram. We note that only the region
between 4.5-7.5 d$^{-1}$ shows any strong peaks, and therefore we zoom
in on this region in Fig.\,\ref{scar}/bottom. The most prominent peak
is the known $\simeq$4 hr variability (indicated as $f_1$) at
6.113$\pm$0.015 d$^{-1}$ (where the error on the frequency is obtained
from the width of the peak). The next two most significant peaks, at
$\simeq$5.1 and $\simeq$7.1 d$^{-1}$, are most likely daily aliases of
the true period. To confirm this, we created an artificial light-curve
of a sine wave with a period of 0.16388875 day (i.e. similar to the
period of Diaz-Trigo et~al. 2009) with the same time-sampling as our
dataset. In the bottom panel of Fig.\ref{scar} we show the resulting
periodogram which indicates that these peaks (labeled as $f_{1,a}$)
are actually daily aliases of $f_1$ and therefore not physical.

More peaks are present that are not reproduced by the periodogram of
our synthetic data (see the bottom panel of Fig\,\ref{scar}). In order
to determine their significance we performed a Monte-Carlo simulation,
using 50,000 random lightcurves that have an identical temporal
  sampling, mean $V$ magnitude and standard deviation as the original
  dataset. For each random lightcurve we obtained the Lomb-Scargle
  periodogram and searched for the period with the highest power. From
  the distribution of the resulting powers we estimated the 3$\sigma$
confidence level and have indicated this in Fig.\,\ref{scar} with a
dashed line. Furthermore, we also created the window function to check
where the most significant peaks will show up due to the sampling of
our data.

Two more peaks, indicated as $f_2$ and $f_3$ (plus their daily aliases
$f_{2,a}$ and $f_{3,a}$) are above the 3$\sigma$ level and do not
correspond to any peak in the window function. The peak at
6.250$\pm$0.015 d$^{-1}$ (i.e. $f_2$) appears to be highly
significant, and we therefore consider this frequency is real. The third
peak ($f_3$), at 5.57$\pm$0.02 d$^{-1}$, is only marginally above our
3$\sigma$ confidence estimate. Although an estimate for the red-noise
component (which our Monte-Carlo simulation does not take into
account) shows that it is not important in this region it could
still lower the significance of this peak enough to make its
identification less certain. However, using the phase dispersion
minimization technique (Stellingwerf 1978), this frequency is also
present at a confidence level $>$3$\sigma$. We therefore tentatively
conclude that this period could also be real, although we do think
that confirmation is still needed.

\section{Spectroscopy}

Our photometric dataset has shown that there are at least 2
significant periods (corresponding to $f_1$ and $f_2$ in
Fig.\,\ref{scar}) present in GR\,Mus, and any physical model will
therefore depend on the interpretation of these periods. Although the
strongest peak has typically been identified as the orbital period
(e.g. Motch et~al. 1987), it is not clear that this is necessarily
correct since it could also correspond to a superhump modulation.  For
example, the optical light could be dominated by the stream-impact
region (and therefore the origin of $f_1$) that has slightly longer
period than the orbital one (which is now $f_2$).  We therefore
obtained 2$\times$4 hrs of phase-resolved spectroscopy to
unambiguously determine the true orbital period. From Barnes
et~al. (2007) we know that a signature of the donor star is present in
the Bowen region and we therefore planned the observations in such a
way that the two candidate orbital periods (corresponding to $f_1$ and
$f_2$) would be in anti-phase with each other during our second set of
spectra.

\subsection{Optical spectrum}

We start by presenting the average spectrum in
Fig.\,\ref{average}. Note that we have not attempted to subtract the
continuum, but have opted to only divide the spectrum by the average
count rate around He\,II $\lambda$4686. The spectrum is dominated by
the emission lines that are typically observed in LMXBs (e.g. Balmer
series, He\,II, Bowen) and, apart from the absence of absorption just
``redward'' of H$\beta$, is identical to that presented by Barnes
et~al. (2007). However, they found several broad unidentified emission
features between 4900 and 5300\AA, which we now interpret as a large
absorption trough from 4900 to 5100\AA. Note that the exact
interpretation of this region will strongly depend on identifying the
precise location of the continuum, which is not possible to model
without a flux standard. We therefore will not attempt here to
identify these features. Finally, at 4530\AA\, there is an unusual
broad feature in our spectrum that was outside the wavelength coverage
of Barnes et~al. (2007). Again, since it will strongly depend on the
exact subtraction of the continuum, it is not possible to identify
this feature and we have therefore labeled it with ``?'' in
Fig.\,\ref{average}.

\begin{figure}\begin{center}
\psfig{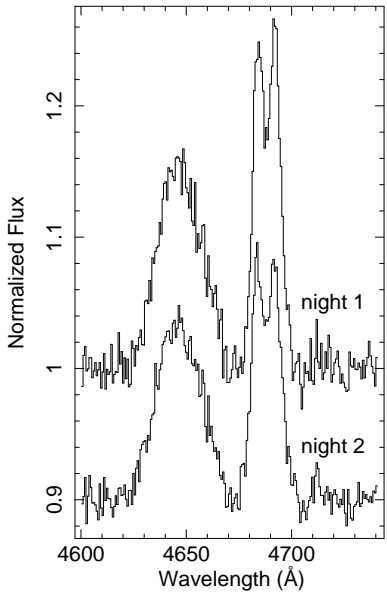}
\caption{Close-up of the average He\,II $\lambda$4686 region of
  GR\,Mus for each individual night, showing that the He\,II profile
  has changed from night to night. Note that the nights have been
  off-set by 0.1 for clarity.
\label{change}}
\end{center}\end{figure}

We also created average spectra for each observing night, and note
that the properties of the major emission lines have changed between
our observations.  In Fig.\,\ref{change} we present the region around
He\,II $\lambda$4686 and the Bowen blend to show that in particular
the strength of He\,II has changed significantly. In order to quantify
these changes we measured for the 3 most prominent emission lines
(i.e. He\,II $\lambda$4686, H$\beta$ and the Bowen region) their
equivalent width (EW) and full width at half maximum (FWHM).

\subsection{Trailed spectra and radial velocities}

Since He\,II $\lambda$4686 is the strongest emission line in our
spectrum we also created a trailed spectrogram of this region. We
first normalized the continuum around He\,II $\lambda$4686 and then
phase-folded the resulting spectra in 20 bins for each night
separately using the ephemeris published in Diaz-Trigo
et~al. (2009). The resulting trails for each night are shown in
Fig.\,\ref{trail}. We have also included the Bowen region in the
trails but no obvious patterns are present there. He\,II $\lambda$4686
on the other hand consists of a clear S-wave with some underlying
structure that could be due to different S-waves. A close inspection
of the bright S-wave shows that it has moved $\simeq$0.05-0.1 orbital
phase between the two nights. Since the blue-to-red crossing for this
bright S-wave occurs around orbital phase 0.7-0.8 it suggests that it
is related to the stream impact region. In Sect.\,4.3 we will explore
this suggestion in more detail when we consider the Doppler maps of
He\,II $\lambda$4686.

\begin{figure}\begin{center}
\psfig{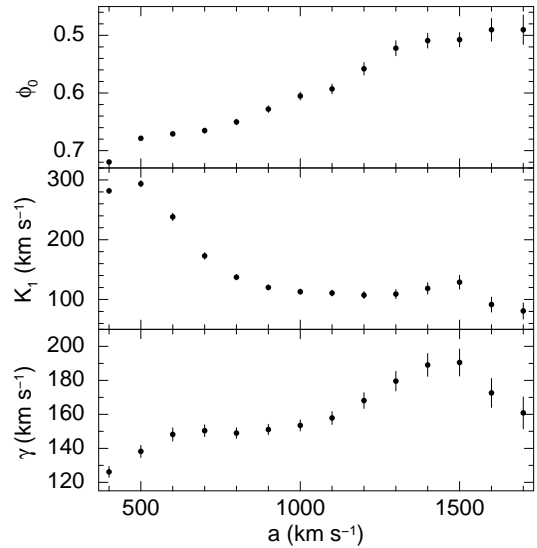}
\caption{Diagnostic diagram for the He\,II $\lambda$4686 emission line
  in GR\,Mus to obtain an estimate for the radial velocity
  semi-amplitude of the compact object ($K_1$), and the systemic
  velocity ($\gamma$) as a function of Gaussian separation $a$.
\label{wings}}
\end{center}\end{figure}

Fig.\,\ref{trail} does suggest the wings of He\,II $\lambda$4686 (in
particular at velocities larger than the semi-amplitude of $\simeq$500
km s$^{-1}$ of the bright S-wave) are not contaminated by any
localized emission sites. We therefore think that the double Gaussian
technique of Schneider \& Young (1980) can be applied in order to estimate the
systemic velocity ($\gamma$) and the velocity semi-amplitude of the
compact object ($K_1$). For the double Gaussian technique we used two
Gaussians with FWHM of 200 km s$^{-1}$, separations ranging between
400 and 1700 km s$^{-1}$, and with steps of 100 km s$^{-1}$, resulting
in the diagnostic diagram shown in Fig.\,\ref{wings}.

\begin{table}\begin{center}
\caption{GR\,Mus Emission line Properties.
\label{stats}}
\begin{tabular}{lccc}
\hline
 & He\,II & H$\beta$ & Bowen\\
\hline
{\bf Night 1} \\
EW   & 3.47$\pm$0.05 & 1.26$\pm$0.05 & 3.92$\pm$0.06\\
FWHM & 899$\pm$19    & 830$\pm$40    & 1655$\pm$53\\
\hline
{\bf Night 2} \\
EW   & 2.70$\pm$0.04 & 0.81$\pm$0.04 & 3.48$\pm$0.05\\
FWHM & 970$\pm$23    & 840$\pm$48    & 1733$\pm$51\\
\hline
\end{tabular}
\end{center}\end{table}

As expected, all Gaussian separations $\le$1200 km s$^{-1}$ are
affected by the bright S-wave that is present in Fig.\,\ref{trail} and
have blue-to-red crossing phase $>$0.5. Furthermore,
Fig.\,\ref{trail} shows that the emission line cannot be distinguished
from the continuum at separations $\ge$1700 km s$^{-1}$
(i.e. $\simeq$26 \AA).  This only leaves separations between 1300-1600
km s$^{-1}$, from which we estimate $\phi_0$=0.51$\pm$0.01 phase,
$K_1$=116$\pm$5 km s$^{-1}$ and $\gamma$=183$\pm$3 km
s$^{-1}$. Interestingly, our estimate for $\phi_0$ is compatible with
emission that is related to the compact object. Furthermore, both our
estimate of $\gamma$ and $K_1$ are consistent with the values obtained
by Barnes et~al. (2007), but due to the higher quality of our dataset
we are able to obtain better constraints.

\subsection{Doppler Maps}

\begin{figure}
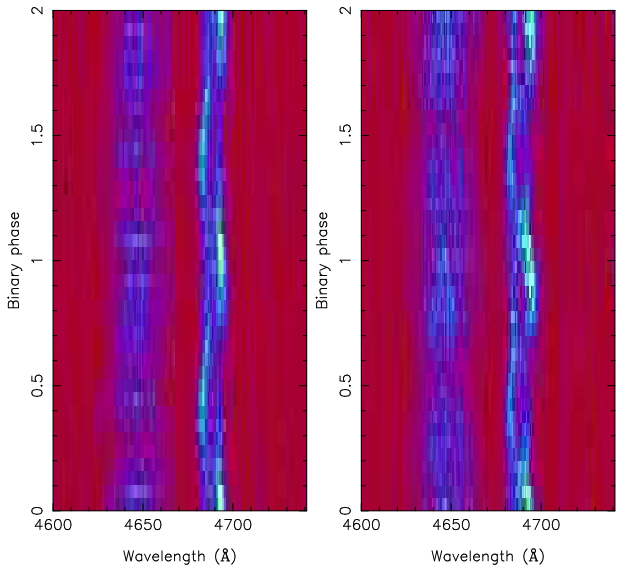
\begin{center}
\parbox{4.0cm}{\psfig{figure=phase_night1.ps,angle=-90,width=4cm}}
\parbox{4.0cm}{\psfig{figure=phase_night2.ps,angle=-90,width=4cm}}
\caption{Trailed spectra of the He\,II $\lambda$4686 and Bowen region
  in GR\,Mus for the first night (left) and second night (right). The
  data has been folded on the ephemeris by Diaz-Trigo et~al. (2009)
  and plotted twice for clarity. A prominent S-wave is visible in the
  He\,II emission line that has slightly changed phasing between the
  two nights.
\label{trail}}
\end{center}\end{figure}

To probe the origin of the different structures in the most prominent
emission lines we used the technique of Doppler tomography (Marsh \&
Horne 1988). This technique also has the advantage that it uses all
spectra simultaneously to enhance features that are too faint in the
individual spectra. This advantage is in particular important for the
Bowen region where the irradiated donor star signature is expected,
since the trail (Fig.\,\ref{trail}) has shown that the individual
spectra do not show anything. Following Barnes et~al. (2007), who also
applied Doppler tomography to detect the irradiated donor star, we
created Doppler maps of the Bowen region for each individual night. We
used the ephemeris by Diaz-Trigo et~al. (2009), the $\gamma$ velocity
determined in Sect.\,4.2, and included the emission lines that are
typically observed in the Bowen region (i.e. N\,III
$\lambda$4640.64/4634.12 and C\,III $\lambda$4647.42/4650.25) to
create the resulting maps shown in Fig.\,\ref{bowen}.

The most important thing to note is that both maps in
Fig.\,\ref{bowen} are dominated by a single compact spot and that is
in the same position, namely situated on the positive $V_y$-axis,
during each night. Since this is the position where a signature of the
irradiated donor star is expected, this not only proves that the
ephemeris of Diaz-Trigo et~al. (2009) is correct, but also that the
true orbital period is 0.1639 days (especially since folding the data
on the other significant period corresponding with $f_2$ in
Fig.\,\ref{scar}, gives two spots in anti-phase to each other).

To test that our assumption for $\gamma$ is correct we also created
Doppler maps of the Bowen region for each night individually by
varying the systemic velocity between 130 and 230 km s$^{-1}$ in steps
of 5 km s$^{-1}$. The spot in both maps was most compact for the
velocities around 180-185 km s$^{-1}$, strongly suggesting that the
$\gamma$ velocity we determined in Sect.\,4.2 is correct. Furthermore
for each night we also produced Doppler maps for a range of $\chi^2$
and note that the spot during the second night is always larger.
These tests strengthened our confidence that the differences in the
spot characteristics are real, i.e. that the spot during the second
night of observations is truly more extended than during the first
night.

Since the Doppler maps of the Bowen region allowed us to determine the
true ephemeris and systemic velocity of GR\,Mus, we can use these
values to create Doppler maps of He\,II $\lambda$4686. The resulting
maps are shown in Fig.\,\ref{hemap}, and confirms what was already
clear from the trail in Fig.\,\ref{trail}. The emission line is
dominated by an extended region around orbital phase 0.7-0.8 (i.e. in
the bottom left quadrant or close to the negative $V_x$-axis) that has
moved between our observing nights. There appears to be a second, much
fainter but more compact spot, that leads the bright region by
$\simeq$0.3 orbital phase (i.e. they are located in the top right
quadrant), that has moved by a similar amount as the bright spot
between the observations. Furthermore, we also created Doppler maps of
H$\beta$, and they show identical behaviour to the He\,II
maps. However, given that the line is much fainter, thereby producing
much lower quality maps, and coupled with the fact that we know from
Barnes et~al. (2007) that the region is complicated due to the
presence of absorption we do not present the resulting maps.

Since we have access to the data by Barnes et~al. (2007) we re-created
their Doppler maps of the He\,II $\lambda$4686 emission line for the
two (consecutive) nights where a full orbital period of GR\,Mus was
observed. In Fig.\,\ref{barnes} we show the result. We note that these
Doppler maps are very similar to that shown in Fig\,\ref{hemap}, with a
bright extended feature in the bottom-left quadrant that moves between
nights, and a fainter more compact spot leading the bright spot and
moving by a similar amount from night to night (although in these
cases only by $\simeq$0.2 orbital phase, placing the spots in the
top-left quadrant). This strongly suggests that the structure in He\,II
$\lambda$4686 is long lived.

\section{Discussion}

\subsection{Accretion disk morphology}

Our 2-week photometric dataset of the LMXB GR\,Mus shows large
morphological changes of its lightcurve from night to night, and
provides further evidence for an evolving accretion disk. Diaz-Trigo
et~al. (2009) suggested that a similar model to that observed in
Her\,X-1, namely a precessing accretion disk that is tilted out of the
plane, could also explain the behavior of GR\,Mus. Although we will
show that this model is actually a good description, we think that the
mechanism for a tilt of the accretion disk in GR\,Mus is different
than is observed in systems like Her\,X-1.

More recent modeling of the 35 day super-orbital period of Her\,X-1
favour a mechanism where the precessing, warped accretion disk is due
to strong irradiation by the central X-ray source (e.g. Wijers \&
Pringle 1999; Ogilvie \& Dubus 2001; Leahy 2002). According to Ogilvie
\& Dubus (2001) such a scenario of radiation-driven warping should
only work for LMXBs with orbital periods $\ge$1 day, and will not work
for systems that have such short orbital periods as GR\,Mus. However, in
the overview of the different mechanisms that could cause periodic
morphological changes in the accretion disks of X-ray binaries by
Kotze \& Charles (2012), they have listed several that could also occur
in GR\,Mus.

\begin{figure}
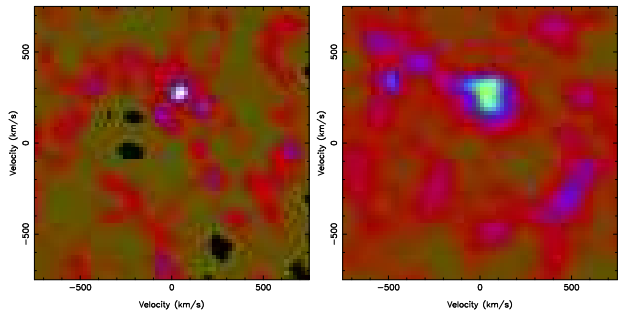
\begin{center}
\parbox{4.0cm}{\psfig{figure=bowen_night1.ps,angle=-90,width=4cm}}
\parbox{4.0cm}{\psfig{figure=bowen_night2.ps,angle=-90,width=4cm}}
\caption{Doppler tomograms of the Bowen region in GR\,Mus for the
  first night (left) and the second night (right). A clear and compact
  spot at the position of the donor star is present during both
  nights, but the exact shape has changed.
\label{bowen}}
\end{center}\end{figure}

One promising effect that could be present in GR\,Mus is the so-called
superhump phenomenon. This effect was first found in a sub-class of
Cataclysmic Variables (CVs), namely the SU\,UMa systems, as the
presence of ``superhumps'' in their lightcurves with periods a few
percent longer than their orbital period (see e.g. Warner 1995 for an
overview).  Although superhumps were originally observed in SU\,UMa
systems, they have also been detected in both black hole LMXBs (see
e.g. O'Donoghue \& Charles 1996; Zurita et~al. 2002) and persistent
neutron star LMXBs such as UW\,CrB (e.g. Hakala et~al. 2009).  They
are thought to be due to a tidal resonance in systems with extreme
mass-ratios (i.e. $q$=$M_{\rm donor}$/$M_{\rm NS}$$\le$0.38) that
initiate the precession of the eccentric accretion disk.  In our
photometric data of GR\,Mus we do find two significant periods that
are only different by a few percent. However, our spectroscopy
unambiguously identifies the longer period (corresponding to $f_1$ in
Fig.\,\ref{scar}) as orbital, and hence the shorter period $f_2$
cannot be attributed to a ``standard'' (prograde) disk precession due
to an eccentric accretion disk (although see Sect.\,5.2 below).

So-called negative superhumps, i.e. systems where a second period
several percent shorter than the orbital one is present, have also
been observed in both CVs and LMXBs (see e.g. Patterson et~al. 1993;
Retter et~al. 2002). The commonly accepted interpretation is that
negative superhumps are due to retrograde precession of a tilted disk
(e.g. Barrett et~al. 1988; Wood \& Burke 2007). Although the exact
details required to produce a tilted disk that has retrograde
precession are still unknown, several promising ideas exist (see
Montgomery 2009a for a more extended discussion). For example, a net
tidal torque by the secondary on an accretion disk that is fully
tilted out of the orbital plane will produce retrograde precession
(Montgomery 2009b), while the differing gas streams over and under the
accretion disk would produce a lift that is strong enough to cause the
tilt (Montgomery 2010).

An important observational parameter is the superhump period deficit
$\epsilon$=($P_{\rm orb}$-$P_{\rm nsh})$/$P_{\rm orb}$. Using our
superhump period and the orbital period by Diaz-Trigo (2009) we
estimate for GR\,Mus that $\epsilon$=0.024$\pm$0.003, which is
comparable to what is observed in other systems with a similar period
(e.g. Montgomery 2009a). Furthermore, Montgomery \& Martin (2010)
predicted that for CVs with system dimensions very similar to GR\,Mus
a mass transfer rate of 8$\times$10$^{-11}$ $M_\odot$ yr$^{-1}$ is
enough to lift the accretion disk out of the plane. Since GR\,Mus has
an average accretion rate at least an order of magnitude larger, this
scenario could explain the behavior of GR\,Mus.

\begin{figure}
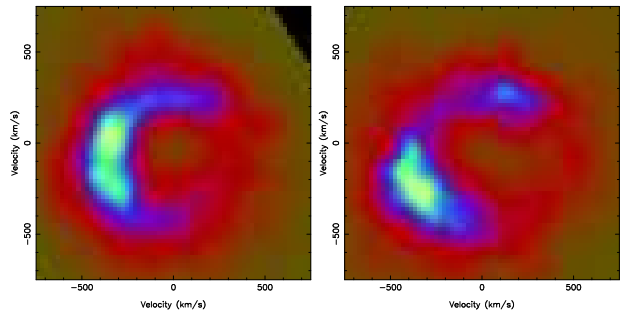
\begin{center}
\parbox{4.0cm}{\psfig{figure=he_night1.ps,angle=-90,width=4cm}}
\parbox{4.0cm}{\psfig{figure=he_night2.ps,angle=-90,width=4cm}}
\caption{Doppler maps of the He\,II $\lambda$4686 emission line for
  the first night (left) and the second night (right), showing a
  bright feature in the bottom-left quadrant of the map that moves
  from night to night.
\label{hemap}}
\end{center}\end{figure}

\begin{figure}
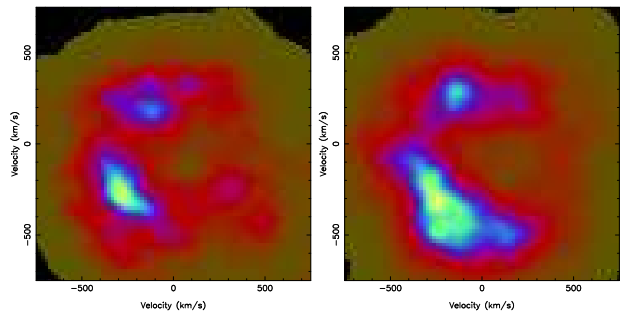
\begin{center}
\parbox{4.0cm}{\psfig{figure=barnes_night1.ps,angle=-90,width=4cm}}
\parbox{4.0cm}{\psfig{figure=barnes_night2.ps,angle=-90,width=4cm}}
\caption{Doppler maps obtained from the dataset by Barnes
  et~al. (2007) of He\,II $\lambda$4686 for the 2 individual nights
  for which a full orbital period was observed. A very similar
  behaviour to that in Fig.\,\ref{hemap} is observed.
\label{barnes}}
\end{center}\end{figure}

The main prediction of this model is that the negative superhumps are
produced by the gas stream that (partly) overflows the disk rim and
instead produces a bright region where it strikes the face of the
accretion disk (Barrett et~al. 1988; Wood \& Burke 2007). The
brightness of this region will vary depending on the amount of
material the (for the observer) visible side of the accretion disk
receives as the secondary orbits (Montgomery 2009a). Furthermore, the
location of this spot would also depend on the precession phase of the
tilted disk. Both these effects are visible in our spectroscopic data.
In the He\,II $\lambda$4686 trail (Fig.\,\ref{trail}) we find that the
brightness of the spot is modulated as a function of orbital phase,
while the corresponding Doppler maps (Figs.\,\ref{hemap} \&
\ref{barnes}) show that the bright region moves from night to
night. The only feature that is not naturally explained by a tilted
disk is the long-lived feature in Figs.\,\ref{hemap}\&\ref{barnes}
that leads the hot-spot by $\simeq$0.2-0.3 orbital phase. Despite this
caveat, we do think that this model gives the best description of our
data, and we propose that GR\,Mus has a disk that is completely tilted
out of the plane along its line of nodes in a similar fashion to
the negative superhumpers observed in CVs.

Although the proposed origin for the tilt is different, the
quantitative model by Diaz-Trigo et~al. (2009), which they based on
one by Gerend \& Boynton (1976), is still correct. Therefore, the main
conclusions by Diaz-Trigo et~al. (2009) to explain most of the X-ray
and optical properties would also still be valid. In this model the
optical emission is a combination of the reprocessed emission in the
accretion disk plus the irradiated surface of the secondary. The
contribution from both components varies depending on the precession
phase of the tilted disk, and gives the exact profile of the optical
lightcurve during each orbital cycle.

\begin{figure}
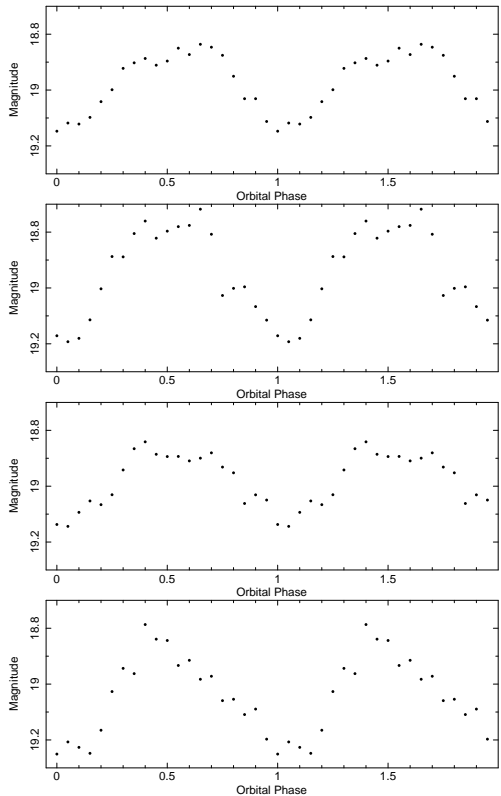
\begin{center}
\psfig{figure=fold_1.ps,angle=-90,width=6.5cm}
\psfig{figure=fold_2.ps,angle=-90,width=6.5cm}
\psfig{figure=fold_3.ps,angle=-90,width=6.5cm}
\psfig{figure=fold_4.ps,angle=-90,width=6.5cm}
\caption{Average photometric lightcurves folded on the ephemeris by
  Diaz-Trigo et~al. (2009) for 4 different phases of the 6.74 day
  precession period (each panel covers 25\% of the precession
  phase). Note that each lightcurve is plotted twice for clarity.
\label{fold}}
\end{center}\end{figure}

From our observations of the negative superhump period of
0.1600$\pm$0.0004 day plus the orbital period of 0.16388875 day by
Diaz-Trigo et~al. (2009) we can calculate a precession period of
$P_{\rm prec}$=($P^{-1}_{\rm orb}$-$P^{-1}_{\rm
  nsh}$)$^{-1}$=6.74$\pm$0.07 days. We note that this period is close
to twice the wait time between the two ``dips'' observed in
Fig.\,\ref{light} (around 9.3 and 12.6 days). This is naively what we
would expect, since the ``dips'' will occur when the accretion disk is
(as close as possible) aligned with our line of sight.  This alignment
will occur twice per precession period, and at exactly half a
precession period separation. Using the precession period and the time
of minimum flux during the second ``dip'', we note that none of our
observations correspond to the time of any other expected
``dip''. Furthermore, we also note that 6.7 days is close to double
the $\sim$60 hr recurrence period of the X-ray dips found by
Diaz-Trigo et~al. (2009). Again this is naively what we should expect,
since half the time the stream would (mainly) flow ``under'' the
accretion disk where it is blocked from our view, and no dipping is
expected during this period. Finally, we divided our photometric
dataset into 4 equal phase bins of the 6.7 day precession period and
phase-folded the resulting lightcurves on the ephemeris of Diaz-Trigo
et~al. (2009). In Fig.\,\ref{fold} we present the resulting
lightcurves, and note that the shape and amplitude of the lightcurves
changes as a function of precession phases. Although each lightcurve
in Fig.\,\ref{fold} covers 25\% of a precession phase it does
illustrate the morphological changes as a function of precession
period.

\begin{figure}\begin{center}
\psfig{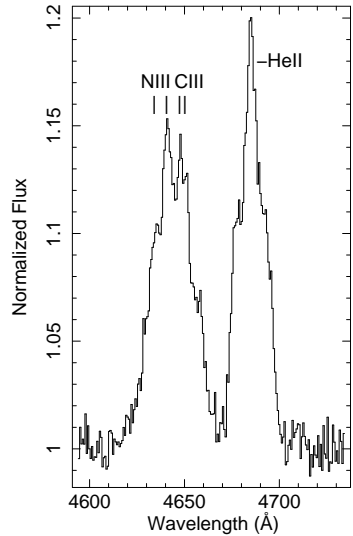}
\caption{Average spectrum of the Bowen region of GR\,Mus in the
  rest-frame of the donor star. We have indicated the dominant
  contributing lines in these regions.
\label{shift}}
\end{center}\end{figure}

An interesting consequence of the stream overflowing the accretion
disk is that there is not necessarily a stream impact region on the
rim of the accretion disk. This bulge is typically used to explain the
X-ray dips that occur around orbital phase $\simeq$0.8 (e.g. Motch
et~al. 1987). Instead, we now have the material overflowing the
accretion disk which can then act as a veil covering the X-ray
producing regions at certain orbital phases, thereby producing the
dips. At other orbital phases the overflowing material would act as
the cold and photo-ionized material between the edge and the inner
accretion disk, thereby producing the absorption features observed at
all orbital and disk precession phases (e.g. Iaria et~al. 2007; Boirin
\& Parmar 2003).

Apart from the change in viewing angle due to the tilt to explain the
global characteristics of the dips (as already outlined by Diaz-Trigo
et~al. 2009), the exact length and strength of each individual dip
will also depend on the density and trajectory of the overflowing
material.  These are components that are much easier to change from
orbit to orbit, instead of changing the structure of the stream-impact
region, and could explain for example the observations by Courvoisier
et~al. (1986).

Finally we note that GR\,Mus is a member of the so-called Atoll
sources, i.e. a group of persistently active LMXBs that have similar
spectral and timing properties in the X-ray regime (Hasinger \& van
der Klis 1989). For those Atoll sources where it is known, their
orbital periods ($\simeq$4 hrs) and accretion rate are also very
similar to that of GR\,Mus, i.e. all the conditions that made the
accretion disk tilt out of the orbital plane in GR\,Mus are also
present in these systems. Coupled with the fact that for  CVS
  there is no correlation between the presence of negative superhumps
  and any of their system parameters (i.e. they can occur for any
  value of $P_{\rm orb}$, $q$, etc; Patterson 1993), it is therefore
possible that a tilted disk is a property of the majority of the Atoll
sources. Since the inclination in other systems is lower, the effect
of a tilted disk will be less obvious (e.g. no dips will be present in
X-rays), but that does not exclude the possibility that its effect is
observable. For example, at the very least the negative superhump
should still be visible in optical photometry and spectroscopy, and
could also be an explanation for the sub-Keplerian features that have
been observed in the emission lines (in particular He\,II
$\lambda$4686) of several LMXBs (e.g. Somero et~al. 2012). A long-term
campaign similar to this one on GR\,Mus should be able to reveal the
presence of a tilted accretion disk in those LMXBs.

\subsection{System parameters}

Barnes et~al. (2007) already concluded that a signature of the
irradiated donor star surface is visible in the Bowen region. The
updated ephemeris by Diaz-Trigo et~al. (2009), together with the clean
and clear spots in the Doppler map of the Bowen region
(Fig\,\ref{bowen}) only strengthens this conclusion. In particular the
Doppler map of the first night shows a very compact spot at the
location very close to where the donor star is expected. From this map
we can estimate a radial velocity semi-amplitude of the irradiated
surface of $K_{\rm em}$=280$\pm$10 km s$^{-1}$, and in
Fig.\,\ref{shift} we have shifted all spectra into the rest-frame of the
donor star to clearly show all the narrow components. We do note that
in both nights the spot is just rightward of the $V_X$-axis (by
$\simeq$30 km s$^{-1}$). This could either be due to a small error in
the phase zero calculations by Diaz-Trigo et~al. (2009) or because the
left hemisphere of the donor star is obscured by the accretion
stream. In any case, this small off-set does not increase the error in
the position of the compact spot. Furthermore, we also note that the
size of the spot is larger during the second night, and this is most
likely due to a lower tilt of the disk blocking less of the donor star
surface than during the first night.

Apart from a more accurate estimate of $K_{\rm em}$, we also have a
better estimate for the radial velocity semi-amplitude of the neutron
star ($K_1$=116$\pm$5 km s$^{-1}$). The fact that both estimates of
$\gamma$ and $\phi_0$ from the diagnostic diagram were similar to the
values that produced the optimal Doppler maps gives us reason to trust
our value for $K_1$. For an ideal LMXB, $K_1$ and $K_{\rm em}$
together with an estimate for the opening angle of the accretion disk,
$\alpha$, would be enough to apply the polynomials of the so-called
$K$-correction by Mu\~noz-Darias et~al. (2005). This allows us to
obtain the radial velocity semi-amplitude at the center of mass of the
donor star ($K_2$). However, since we now know that GR\,Mus harbours a
tilted accretion disk, it is not possible to apply a ``simple''
$K$-correction. Instead we can explore what the effect on the system
parameters are, assuming that the opening angle in GR\,Mus is a
combination of the traditional disk opening angle plus the tilt of the
accretion disk.

De Jong et~al. (1996) estimated an opening angle of 12$^\circ$ for
GR\,Mus, which is close to the 9-13$^{\circ}$ by Motch
et~al. (1987). Furthermore, modeling by Motch et~al. (1987) showed
that the bulge region has an azimuthal height of 17-25$^{\circ}$,
which can change to heights $\le$10$^{\circ}$ (Smale \& Wachter
1999). Obviously, the true final opening angle will depend on the
exact tilt of the disk at the time of our spectroscopic observations.
We can still derive an estimate for the $K$-correction using the
polynomials by Mu\~noz-Darias (2005) for opening angles between
8$^{\circ}$ and 14$^{\circ}$. Note that due to the compactness of the
spot in Fig.\,\ref{bowen}(left) we do think larger opening angles are
more likely.  These $K$-corrections lead to a range for the mass ratio
of $q$=0.29-0.35, with $q$ increasing for larger opening angles.

Interestingly, for this range of $q$ it is possible to have an
eccentric accretion disk due to tidal resonance (see Sect.\,5.1;
Whitehurst 1988). If GR\,Mus does have an eccentric accretion disk our
photometric observations should have shown the presence of positive
superhumps. Interestingly, there is the marginal significant peak
$f_3$ in Fig.\,\ref{scar} at 5.57$\pm$0.02 d$^{-1}$
(=0.1795$\pm$0.0007 day). Although uncommon, both negative and
positive superhumps have been observed in both CVs and LMXBs (Retter
et~al. 2002; Olech et~al. 2009 and references therein). Furthermore,
Patterson et~al. (2005) showed that an empirical relationship exists
between the superhump period excess ($\epsilon$=($P_{\rm sh}$-$P_{\rm
  orb}$)/$P_{\rm orb}$=0.18$q$+0.29$q^2$) and the mass ratio. Assuming
that peak $f_3$ is truly a positive superhump (leading to
$\epsilon$=0.095$\pm$0.005) we can estimate a mass ratio of
$q$=0.33-0.36. This is remarkably close to what we obtain from the
$K$-correction.

Retter et~al. (2002) found a relation between orbital period and ratio
between the period excess and deficit of the two superhumps. Olech
et~al. (2009) refined this relation by including more CVs that showed
both negative and positive superhumps to
$\phi$=$\epsilon_-$/$\epsilon_+$=0.318$\times$log$P_{\rm orb}$-0.161,
where $\phi$ is negative and $P_{\rm orb}$ is in days. Although
$\phi$$\simeq$-0.25$\pm$0.006 for GR\,Mus, and is therefore relatively
far removed from the relation of Olech et~al. (2009), the scatter in
the few points that form this relation is still large. Given that the
$\phi$ for GR\,Mus is not too different compared with other systems
that have a similar orbital period, we think it provides another
indication that the positive superhump could be real.

We can further constrain the mass of GR\,Mus if we assume that it
  truly harbours an eccentric precessing accretion disk, and that the
  relationship between $\epsilon_+$ and $q$ by Patterson et~al. (2005)
  is true (thereby providing the strongest constraint on $q$ of
  0.33-0.36). First of all, from the fact that GR\,Mus does not show
any eclipses we can obtain an upper-limit on the inclination of
$\le$73$^{\circ}$ (Paczynski 1974). However, the true inclination will
depend on both the tilt of the accretion disk and its opening angle,
and could be 12-15$^{\circ}$ lower. Under the assumption that the
inclination is close to $i$=73$^{\circ}$, we obtain a mass of the
neutron star of 1.3$\pm$0.2$M_\odot$, which is very close to the
canonical mass of a neutron star. However, if we only assume that
$i$$\le$73$^{\circ}$, we obtain a lower-limit on the neutron star mass
of $\ge$1.2$M_\odot$ (95\% confidence level).

Finally, taking into account that modeling by Motch et~al. (1987)
gives a lower limit on the inclination of $i$$\ge$65$^{\circ}$, we
obtain a maximum mass on the neutron star of $\le$1.8$M_\odot$ (95\%
confidence). This suggests that the largest uncertainty to obtain the
mass of GR\,Mus is the inclination, which can hopefully be better
constrained with detailed modeling of our lightcurves.

\section{Conclusions}

We have detected the presence of superhumps in our 14 nights of
photometric observations of the LMXB GR\,Mus. Together with 2$\times$4
hrs of spectroscopic observations we identify them as negative
superhumps, which are due to an accretion disk that is completely
tilted out of the orbital plane along its line of nodes and
shows retrograde precession. This not only gives strong support to the
model proposed by Diaz-Trigo et~al. (2009), but also provides a
theoretical framework for further modeling of the optical and X-ray
lightcurves of GR\,Mus. Due to its high inclination we think that
GR\,Mus is an intriguing candidate to further our understanding of
negative superhumps. In particular, the existing X-ray observations
provide an excellent way to probe the effect of a tilted disk on our
view of the inner accretion disk, while the photometry does the same
for the outer accretion disk.

Furthermore, we also find marginal evidence for the presence of an
eccentric accretion disk in the form of positive superhumps. If true,
this provides strong constraints on the mass ratio, and together with our
constraints on $K_1$ and modeling of the inclination by Motch
et~al. (1987) gives a mass of the neutron star of 1.2$\le$$M_{\rm
  NS}$/$M_{\odot}$$\le$1.8 (95\% confidence level). Interestingly, we
also note that the modeling of the opening angle of the accretion disk
by Motch et~al. (1987) give $\alpha$=9-13$^{\circ}$. This angle is
very similar to that needed to provide the proper $K$-correction
(namely 12-15$^{\circ}$) using the polynomials by Mu\~oz-Darias
et~al. (2005). This does suggest that we have obtained a
consistent set of system-parameters, with the largest
uncertainty remaining the inclination of GR\,Mus.

\section*{Acknowledgments}
This work is based on data collected at the South African Astronomical
Observatory and the European Southern Observatory, Paranal, Chile
[Obs. Id. 089.D-0274]. We acknowledge the use of PAMELA and MOLLY
which were developed by T.R.  Marsh, and the use of the on-line atomic
line list at http://www.pa.uky.edu/$\sim$peter/atomic.  RC
acknowledges a Ramon y Cajal fellowship (RYC-2007-01046) and a Marie
Curie European Reintegration Grant (PERG04-GA-2008-239142). RC and JC
acknowledge support by the Spanish Ministry of Science and Innovation
(MICINN) under the grant AYA 2010-18080. This program is also
partially funded by the Spanish MICINN under the consolider-ingenio
2010 program grant CSD 2006-00070.

\bsp

\label{lastpage}

\end{document}